\documentclass[sigconf]{acmart}
\AtBeginDocument{%
  \providecommand\BibTeX{{%
    \normalfont B\kern-0.5em{\scshape i\kern-0.25em b}\kern-0.8em\TeX}}}
\usepackage{amsfonts}

\setcopyright{acmcopyright}
\copyrightyear{2023}
\acmYear{2023}
\acmDOI{XXXXXXX.XXXXXXX}

\acmConference[KDD'23]{Workshop on Decision Intelligence and Analytics for Online Marketplaces}{August 06--10,
  2023}{Long Beach, CA}
\acmPrice{15.00}
\acmISBN{978-1-4503-XXXX-X/18/06}

\begin{document}
\title{Bid Optimization for Offsite Display Ad Campaigns on eCommerce}

\author{Hangjian Li}
\email{hangjian.li@walmart.com}
\affiliation{%
  \institution{Walmart Global Tech}
  \city{Bellevue}
  \state{Washington}
  \country{USA}
  \postcode{98004}  
}

\author{Dong Xu}
\email{dong.xu@walmart.com}
\affiliation{%
  \institution{Walmart Global Tech}
  \city{Sunnyvale} 
  \state{California}
  \country{USA}
  \postcode{94086}
  }

\author{Konstantin Shmakov}
\email{konstantin.shmakov@walmart.com}
\affiliation{%
  \institution{Walmart Global Tech}
  \city{Sunnyvale}
  \state{California}
    \postcode{94086}
  \country{USA}}

\author{Kuang-Chih Lee}
\email{kuangchih.lee@walmart.com}
\affiliation{%
  \institution{Walmart Global Tech}
  \city{Sunnyvale}
  \state{California}
  \country{USA}
  \postcode{94086}
}
\author{Wei Shen}
\email{wei.shen@walmart.com}
\affiliation{%
  \institution{Walmart Global Tech}
  \city{Sunnyvale}
  \state{California}
  \country{USA}
  \postcode{94086}
}


\renewcommand{\shortauthors}{Li, et al.}
\newcommand{\thickhline}{\hlineB{4}}
\newcommand{\NA}{---}
\newcommand{\todo}[1]{\textcolor{red}{#1}}
\begin{abstract}
  Online retailers often use third-party demand-side-platforms (DSPs) to conduct offsite advertising and reach shoppers across the Internet on behalf of their advertisers. The process involves the retailer participating in instant auctions with real-time bidding for each ad slot of their interest. In this paper, we introduce a bid optimization system that leverages the dimensional bidding function provided by most well-known DSPs for Walmart offsite display ad campaigns. The system starts by automatically searching for the optimal segmentation of the ad requests space based on their characteristics (e.g. geo location, device type). Then, it assesses the quality of impressions observed from each dimension based on revenue signals driven by the campaign effect. During the campaign, the system iteratively approximates the bid landscape based on the data observed and calculates the bid adjustments for each dimension. Finally, a higher bid adjustment factor is applied to dimensions with potentially higher revenue over ad spend (ROAS), and vice versa. The initial A/B test results of the proposed optimization system has shown its effectiveness of increasing the ROAS and conversion rate while reducing the effective cost per mille for ad serving. 
\end{abstract}





\maketitle

\section{Introduction}
Programmatic advertising has been one of the greatest innovations of eCommerce over a long time. By replacing much of the manual work of managing ad inventory, negotiating prices, discovering targeted users, purchasing ad space, etc. between an advertiser and a publisher with automated systems, programmatic advertising makes the transaction much easier, secure, and efficient. It achieves this by building an open marketplace, or sometimes called an ``ad exchange'', that enables all advertisers and publishers to trade with each other online in real-time.  Sitting at the core of the online open marketplace is what's called a \textit{real-time bidding} system, which enables an ad opportunity to be sold via a real-time auction among all advertisers. 

\begin{figure}
    \centering
    \includegraphics[width=\linewidth]{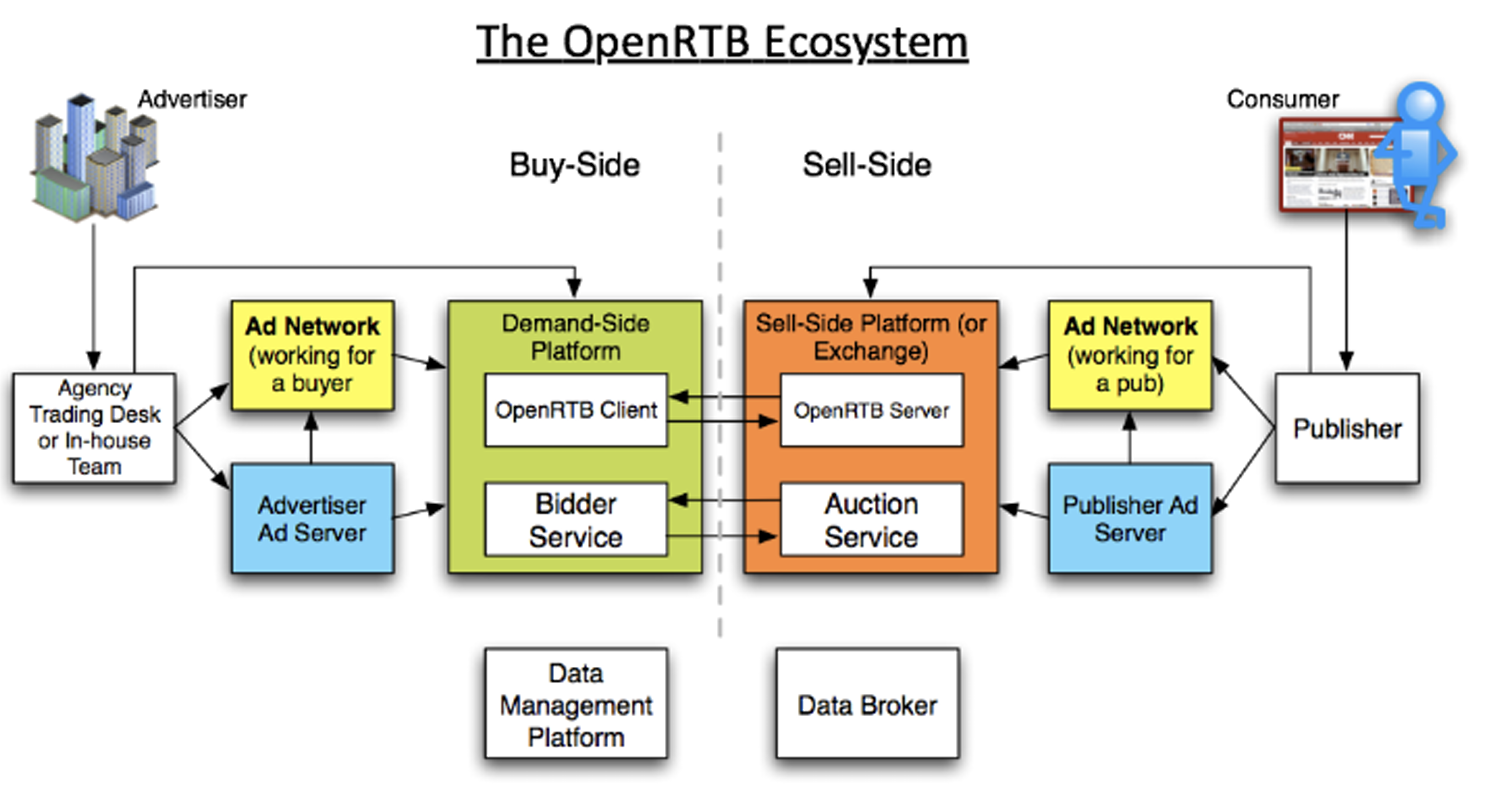}
    \caption{Parties involved in the online advertising system (Source: OpenRTB 2.3 \cite{openRTB})}.
    \label{fig:openrtb}
\end{figure}
The real online advertising systems are more complicated than what's described above, and involves other parties such as demand-side-platforms (DSPs) and supply-side-platforms (SSPs) to participate in RTB auctions on behalf of advertisers and publishers, respectively --- see Figure ~\ref{fig:openrtb}. DSPs have been created to give ad inventory buyers control over their programmatic campaign strategies such as bidding prices for ad auctions, audience targeting preferences, budgeting pacing and throttling, etc, which are executed through the real-time bidding auction \cite{kosorin2016introduction}. Popular DSPs include Facebook Ads Manager, Google DoubleClick, The Trade Desk, AppNexus, etc. Online retailers have been a major client of DSPs as they have huge demand for online ads. As the eCommerce business picks up the pace, a number of eCommerce companies are also investing heavily into creating their own in-house DSPs either through collaboration with the existing leading DSPs or by themselves. In-house DSPs offer obvious benefits such as better integration, more customized functionalities, and overall more control on advertising strategies. Players in this area include but not limited to Amazon, Walmart, Apple, Alibaba, eBay, etc. In this work, we focus on the bid adjustment tool offered by most DSPs and discuss ways of optimizing the adjustment factors towards ROAS from the perspective of online retailers.

\section{Background and Literature}\label{subsec: offsite}
Bid optimization for display advertising in online RTB has been studied extensively over the years in both academia and the private sector. RTB auctions was initially developed for sponsored search to sell search impressions by posting a bid for user query keywords \cite{10.1257/aer.97.1.242, mcafee2011design}. Later on, display advertising 
started to employ RTB auctions to sell an ad impression whenever it is generated by a user's visit \cite{10.1145/2501040.2501980}. The research topics for display advertising can be divided into the demand-side and the supply-side. The key issues faced on the supply side involves inventory prioritization \cite{doi:10.1287/mnsc.2014.2017}, allocation \cite{abolhassani2022online}, forecasting \cite{zhang2014optimal}, etc., which are all important topics but are not covered in today's discussion. The most important topic on the demand-side is the design and implementation of effective bidding algorithms. It is usually formulated as a constrained optimization problem where restrictions on the bid price, budget, pacing, etc. are enforced and we want to maximize metrics such as total impressions won (reach), total clicks, total conversions, ROAS, etc. The optimization can be formulated at the impression-level based on a linear
programming prime-dual problem, and have a bidding
algorithm that enables fine-grained impression valuation, and
adjusts value-based bid according to real-time constraints \cite{chen2011real}. Another way is to employ ``multiplicative bidding'' where the effective bidding price is equal to the product of a base bid and multiple bid adjustments that are determined by the features of the ad request \cite{google, 10.1145/2600057.2602874}. This allows advertisers to optimize towards their metrics of interest by expressing relative valuations across a fixed set of feature combinations defined by the DSP. In this paper, we also follow this route and investigate how we can optimize through bid adjustments. Since the design of bidding strategies typical depends heavily on the distribution of future winning bid values, an accurate \textit{bid landscape} forecast model is critical. Wang et. al \cite{wang2017display} proposed different bid landscape models while others such as Lang et. al \cite{10.1145/2187836.2187887} discussed optimal bidding strategies when inaccurate bid landscape predictions are present. 


Many online retail companies do not have the resource or the need to build an in-house DSP that keeps track advertising opportunities across the Internet (offsite). For most ad campaigns, the objectives are usually maximizing the number of impressions delivered (customer reach), clicks per impression (CTR), conversion rate (CVR), and revenue (e.g., ROAS) by bidding strategically on each impression opportunity under a budget constraint. However, there is a critical gap of information collected by 3P DSP and online retailer themselves: DSPs collect the information about online ad auction and bidding environment such as ad spend, clicks, win rate, pacing, etc. for multiple retail platforms, while retailers mainly owns the conversion and revenue signals of products sold on their own platform, such as conversion amount, order amount, order value, etc. Due to confidentiality or propriety restrictions, both sides will only share limited information they collected, making it difficult for bidding optimization. For example, DSPs cannot automatically optimize bid prices towards maximizing revenue as they do not have revenue data. As a result, the campaign managers must manually adjust the bid prices using their own discretion if they want to maximize revenue-related metrics such as ROAS. This is both labor-intensive and inaccurate because the retail campaign managers may not have the detailed insights into the online auction market. 

Despite the fact that most well-known DSPs nowadays provides optimization features towards CTR, we found that the correlation between the click-based metrics and revenue is low for display campaigns due to their delayed effect on conversion. In general, online retailers face two major difficulties when trying to optimize bids towards ROAS. First, bid optimization w.r.t revenue relies on bid landscape prediction, a model which returns the probability of winning an auction given a bid price. Retailers, on the other hand, only possesses the information of the impression they have won, making it very difficult to estimate the bid landscape (in some cases, DSPs may choose to share win rate of the retailer at certain aggregated granularity but it is usually not sufficient for bid landscape modeling). Second, 3P DSPs control the pacing and throttling systems that are completely unexposed to retailers. As a result, the effect of any bid adjustment by online retailer is confounded by the intervention of such systems. And the patterns directly observed by retailers from their censored data of bid price, cost, revenue, etc. are often unstable if not misleading. 

\subsection{Dimensional Bidding}
Despite not being directly supported by 3P DSPs, revenue-based optimization may be approximated using a \textit{dimensional bidding} approach. For example, Google Ad Manager \cite{google} allows campaign managers to apply bid adjustments for ad requests from different segments. Other DSPs also offer similar features. This service is usually provided through an API. 

\begin{figure}
    \centering
    \includegraphics[width=0.7\linewidth]{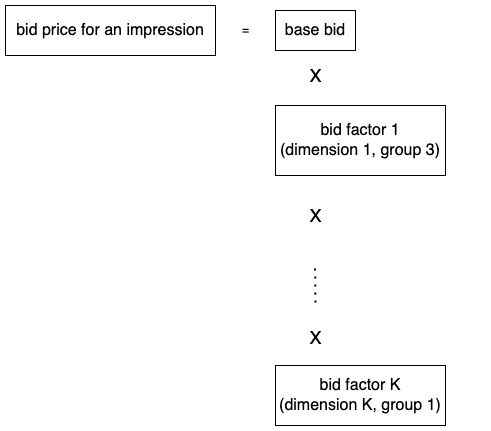}
    \caption{Dimension bidding with group bid adjustments. The bid price for each ad request is determined by a base bid and several bid adjustment factors. The value of each bid adjustment factor can be set by retailer campaign managers.}
    \label{fig:dim_bid}
\end{figure}

The dimensional bidding API allows campaign managers to define \textit{bid dimensions} of the impression space and apply different bid adjustments to the impressions in different dimensions. A dimension is a targeting item that one can target, block, or apply bid adjustments to. For example, they can be defined on domain, ad format, geolocation, time, device type, etc. The final bid price will be the product of a base bid and several bid adjustment factors, as illustrated in Figure \ref{fig:dim_bid}. To use this tool, retail campaign managers need to first define the bid dimensions. And the optimal way to do that depends on how these adjustments would affect the campaign spend and performance. Thus, we need to understand and model the relationship among these key factors.

Our main contributions include the following:
\begin{enumerate}
    \item We proposed a new automated bid optimization system to maximize the ROAS of display ad campaigns run on 3P DSPs.
    \item We proposed a novel way of defining and evaluating the best-performing bid dimensions based on historical data. 
    \item In approximating the bid landscape with censored data, we proposed a 
    marginalized approach to reduce model complexity without compromising flexibility. 
    \item We demonstrated through A/B online test that our optimization system can achieve higher ROAS performance. 
\end{enumerate}

The rest of the paper is organized as follows: In Section ~\ref{sec:method} we first talk about an simple optimization framework based on disjoint impression segments. Then we introduce our main approach based on overlapping dimensions. We also cover our bid landscape forecast model in this section. In Section ~\ref{sec:dimension} we discuss different ways of selecting dimensions and creating impression segments based on each dimension. Various ways of evaluating the generated dimensions and segments are also discussed. In Section ~\ref{sec:results} we cover our online A/B test results for the proposed bid optimization method. We conclude the paper and discuss the future work in Section ~\ref{sec:summary}.

\section{Bid Optimization Methods}\label{sec:method}
In this paper we assume the ad campaign budget is given for each line item, and the objective is to maximize ROAS for each line item. Equivalently, we want to maximize the total revenue for each line item under the budget constraint. Because we cannot accurately model the bid landscape for reasons explained in Section \ref{subsec: offsite}, we will rely on a few assumptions to establish our optimization framework.   

Before laying out the assumptions, let’s define a few quantities and notations. Let $b_0$ denote the base bid price. The cost per (mille) impression (CPM) is defined as the total ad spend divided by ad volume (multiply by one thousand). The revenue per (CPM) impression (RPM) is defined as the total attributed revenue divided by ad volume (multiply by one thousand). Note that the amount of attributed revenue depends on the size of the attribution window. 

The basic idea of our bid optimization is to divide the impression population into sub-populations, i.e., dimensions, so that the behaviors of these sub-populations are easier to track and be exploited. Retailers usually have complete control over which dimensions to use and how sub-populations should be defined for each dimension. We start by making three assumptions on the impressions from each dimension:

\begin{enumerate}
    \item The CPM is an increasing function of the bid factor. 
    \item The number of winning auctions (or impressions we get) is an increasing function of the bid factor.
    \item RPM is independent from the bid price and relatively stable for a short period of time (1-2 days). 
\end{enumerate}

Display campaigns’ ad cost are determined by media cost (measured by cost-per-mille (CPM)) plus some additional fees (usually they are constant or a fixed percentage). Because the bid price for impressions under the same dimension is equal to the base bid times bid adjustment, CPM should increase if the bid adjustment increases. The second assumption is also intuitive as the number of winning actions should also be proportional to the bid price. The last assumption on RPM says that if the dimensions are chosen based on RPM, the average revenue we can get from each dimension should not depend on how much we paid for them. For example, if an ad slot on Yahoo.com during certain prime hour in a zip code region usually gets $x$ amount of attributed revenue during a normal day. Despite the fluctuation in the bid landscape, the revenue we get from winning this ad slot won’t change much regardless of the actual ad spend. In practice, the validity of this assumption depends on the quality of the sub-populations created, and we will discuss more about this in Section ~\ref{sec: quality}. The following types of data (at impression level) are used throughout the rest of this paper: 
\begin{enumerate}
    \item Daily attributed revenue (14 or 30-day attribution window) for each impression served;
    \item Impression cost (equal to bid price under first price auction);
    \item Impression attributes (time, geo, publisher, platform, etc.); 
    \item campaign metadata (budget, start and end dates, etc.)
\end{enumerate}

\subsection{Optimize with Disjoint Bid Dimensions}\label{sec: disjoint}
Under the previous assumptions, we first came up with a straightforward optimization framework by defining multiple disjoint dimensions and apply unique bid adjustments to each dimension. Let’s define some notations first. For dimension $i$, we define the following quantities:
\begin{itemize}
    \item Cost per (mille) impression: $\mathrm{CPM}_i$.
    \item Revenue per (mille) impression: $RPM_i$.
    \item Number of won impressions: $n_i$.
    \item Total ads spend: $S_i$.
    \item Total revenue: $R_i$.
    \item Bid (adjustment) factor: $f_i$.
\end{itemize}

\begin{table}[]
    \centering
    \begin{tabular}{c|c | c| c | c}
    \hline
        & $\text{Geo}_1$ & $\text{Geo}_2$ & $\text{Geo}_3$ & total   \\
    \hline
        $\text{website}_1$ & 1  & 2 & 3 & 6\\ 
    \hline
         $\text{website}_2$ & 12& 1& 6& 19\\ 
    \hline 
         $\text{website}_3$ & 7& 10& 9&26\\ 
    \hline
         total &20& 13& 18& 51\\ 
         \hline
    \end{tabular}
    \caption{Example of impression volume in each dimension. Two dimensions: website \& geo location are used in this example. The total number of impressions is equal to the sum of all individual cells in the table.}
    \label{tab:simple_table}
\end{table}

\begin{table}[]
    \centering
    \begin{tabular}{c|c | c| c | c}
    \hline
        & $\text{Geo}_1$ & $\text{Geo}_2$ & $\text{Geo}_3$ & total   \\
    \hline
        $\text{website}_1$ & $n_1=\beta_1 f_1$ & $n_2 = \beta_2f_2$ &$ n_3 = \beta_3f_3$ & $\sum_{i=1}^3 \beta_i f_i$\\ 
    \hline
         $\text{website}_2$ &$n_4=\beta_4 f_4$ & $n_5 = \beta_5f_5$ &$ n_6 = \beta_6f_6$ & $\sum_{i=4}^6 \beta_i f_i$\\ 
    \hline 
         $\text{website}_3$ & $n_7=\beta_7 f_7$ & $n_8 = \beta_8f_8$ &$ n_9 = \beta_9f_9$ & $\sum_{i=7}^9 \beta_i f_i$\\
    \hline
         total & $\ldots$& $\ldots$& $\ldots$& $\sum_{i=1}^9\beta_i f_i$ \\
    \hline
    \end{tabular}
    \caption{Simple linear model for impression volumes. Each cell corresponds to a dimension with a unique bid factor.}
    \label{tab:simple_table2}
\end{table}
	
Let us also define a few addition campaign variables such as the total budget $B$ and the flight duration $T$. Then, the optimization problem needs to be solved is in Eq. ~\eqref{eq:simp_obj}. 

\begin{equation}\label{eq:simp_obj}
\begin{aligned}
    \max_{f_i}\ &\sum_{i=1}^m RPM_i\cdot g(f_i) \\
    \textrm{s.t.}\ & \sum_{i=1}^m S_i\leq B 
\end{aligned}
\end{equation}
where $n_i = g(f_i)$ is a model of our choice for impression volume from dimension $i$. Without the loss of generality, we assume base bid $b_0 = 1$. Taking a simple linear model as an example, in Eq. ~\eqref{eq: linear}, we postulate that the total number of impressions we can win from dimension $i$ increases linearly with the corresponding bid adjustment factor. 
\begin{equation}\label{eq: linear}
    n_i = g(f_i) = \beta_i\cdot f_i
\end{equation}
Table ~\eqref{tab:simple_table} and \eqref{tab:simple_table2} give a concrete example of modeling the impression volumes using linear functions when there are 9 disjoint dimensions defined by geo location and websites. In the example, both geo and website define 3 groups but in practice the number of groups do not need to be the same. 

Under the first price auction setting, $CPM_i \approx b_0\cdot f_i = f_i$. From the assumptions above one can derive the following relations:
\begin{align}
    S_i &= n_i\cdot CPM_i = \beta_i\cdot f_i^2\\
    R_i &= RPM_i\cdot n_i = RPM_i\cdot \beta_i\cdot f_i = RPM_i\cdot\sqrt{\beta_i\cdot S_i}.
\end{align}
and the optimization problem becomes:
\begin{equation}\label{eq: easy}
    \begin{aligned}
        \max_{f_i}\ &\sum_{i=1}^m RPM_i\cdot\beta_i\cdot f_i\\
        \textrm{s.t.}\ & \sum_{i=1}^m f_i^2\cdot \beta_i \leq B.
    \end{aligned}
\end{equation}
Solving Eq.~\eqref{eq: easy} yields the optimal bid adjustment factors:
\begin{equation}\label{eq: optimal1}
    \hat f_i = \sqrt{\frac{RPM_i^2}{\sum_{j}^mRPM_j^2\cdot\beta_j}\cdot B}
\end{equation}
Suppose we divide the impression space into disjoint dimensions and assign a unique bid adjustments to each, Eq.~\eqref{eq: optimal1} shows that the optimal bid adjustment for dimensions $i$ should be proportional to $RPM_i$ and equal to a fraction of budget, given by $m$ linear impression volume models. 

However, this framework has two main drawbacks. First, it might suffer from the \textit{curse of dimensionality}. Instead of defining the bid adjustment as a product, it has a unique factor for each segment. In practice, if the number of segments is larger, we cannot reliably estimate $RPM_i$, the volume model $g\left(f_i\right)$,or the cost model \cite{Bellman1957}. Second, the simple approach is not scalable as the number of dimensions and groups increase. The number of parameters is of order $\Theta\left(I^K\right)$ where $K$ is the number of dimensions and $I$ is the number of groups within each dimension.

\subsection{Optimize with Overlapping Bid Dimensions}
\begin{table}[]
    \centering
      \renewcommand{\arraystretch}{3}
      \resizebox{\linewidth}{!}{
    \begin{tabular}[width=\linewidth]{c|c | c| c |c }
    \hline
        & $\text{Geo}_1 (f_1^2)$ & $\text{Geo}_2 (f_2^2)$ & $\text{Geo}_3(f_3^2)$ & total  \\
    \hline
        $\text{website}_1 (f_1^1)$ & \NA&\NA&\NA& $n_1^1 = g\left(f_1^1, f_{[1,2,3]}^2\right)$\\ 
    \hline
         $\text{website}_2 (f_2^1)$ & \NA&\NA&\NA&$n_2^1 = g\left(f_2^1, f_{[1,2,3]}^2\right)$\\
    \hline 
         $\text{website}_3 (f_3^1)$ & \NA&\NA&\NA &$n_3^1 = g\left(f_3^1, f_{[1,2,3]}^2\right)$\\
    \hline
    total &$n_1^2 = g\left(f_1^2, f_{[1,2,3]}^1\right)$&$n_2^2 = g\left(f_2^2, f_{[1,2,3]}^1\right)$&$n_3^2 = g\left(f_3^2, f_{[1,2,3]}^1\right)$& $=\sum_{i=1}^3 n_i^1 = \sum_{i=1}^3 n_i^2$\\
    \hline
    \end{tabular}}
    \caption{Impression volume by dimension and group under the over-lapping dimension setup. Bid dimensions and groups are now defined marginally and the adjustment factors can interact with each other. The total volume is equal to either the sum of website group volumes or the sum of geo group volumes.}
    \label{tab:overlap}
\end{table}
To address the drawbacks above, we came up with a more general framework. Let’s first look at an example. Similar to Table ~\eqref{tab:simple_table}, in Table ~\eqref{tab:overlap}, we split the impression space based on two dimensions: site and geo, where 3 groups are created based on geo and 3 based on site. However, this time instead of assigning a unique bid factor to each segment, we assign one to each group. Concretely, for group $j$ of dimension $i$, we denote its bid factor by $f_j^i$, as shown next to the group labels. Instead of having $3\times3 = 9$ bid factors, we end up with only $3 + 3 = 6$ bid factors. This is a significant reduction in the number of parameters we need to keep track of. However, because the bid price for impressions will be determined by two bid factors instead of one, the modeling for the impression volume and cost will get slightly more complicated. Concretely, the bid price under the new setting is given by 
\begin{equation}
    b = b_0 \cdot f_{w}^1 \cdot f_{z}^2
\end{equation}
where $w$ is the website group the impression belongs to and $z$ is its geo group index. Comparing to the method described in Section ~\ref{sec: disjoint}, this method makes an implicit assumption on the bid factors being decomposable (suppose there are only geo and site dimensions):
\begin{equation}
    f_l = f_w^1\cdot f_z^2
\end{equation}
where $l$ is one of the disjoint dimensions that corresponds to site $w$ and geo $z$. The number of parameters under the new setup reduces from $\Theta(I^K)$ to $\Theta(I\cdot K)$.

Yet we still need to define the relationship among key variables such as revenue, spend, bid price, etc. Without the loss of generality, let there be K dimensions and each dimension have I groups (in practice the number of groups can be different). We still define the dimensional level spend and revenue as in Section ~\ref{sec: disjoint} except now the sum of group ad spend along any dimension\ i is equal to the total ad spend:
\begin{align}
    S^i &= \sum_{j=1}^I S_j^i = B \\
    \sum_{i=1}^K S^i &= K\cdot B.
\end{align}
The same rule applies to revenue and impression volumes. 
Let’s take an impression that belongs to $\text{website}_1$ and $\text{Geo}_3$ as an example. The bid price for this impression would be equal to 
$b=b_0\times f_2^1\times f_3^2$ following the notation in Table ~\ref{tab:overlap}. Because a group from one dimension now overlaps from any group from another dimension, the impression volume $n_i^k$, revenue $RPM_i^k$, as well as the cost $CPM_i^k$ can be estimated based on more impressions: an entire row or column in Table ~\ref{tab:overlap} versus a single cell in Table ~\ref{tab:simple_table2}. The total revenue and cost can be represented as 
\begin{align}\label{eq:total_rev}
    R &= \frac{1}{K}\sum_{k=1}^K \sum_{i=1}^{I_k}n_i^k\cdot RPM_i^k\\
    S &= \frac{1}{K}\sum_{k=1}^K \sum_{i=1}^{I_k}n_i^k\cdot CPM_i^k
\end{align}

To maximize the total revenue $R$ in E.q ~\eqref{eq:total_rev} w.r.t. the bid factors $f_i^k$ such that $S\leq B$, we need to make a few assumptions on the relationship between impression volume $n_i^k$ and cost $CPM_i^k$ vs. the bid factors $f_i^k$. In other words, we need to model the number of impressions we can win and their average cost w.r.t. the bid adjustment factors. This is similar to the bid landscape models that DSPs often build for campaign optimization, and they are usually refreshed with high frequency (every minute or hour) based on the most recent data. Since retailers only get censored impression log from 3P DSPs with some delay (DSPs often share logs on a daily basis), we will rely on a less accurate bid landscape approximation. 

\subsection{Bid Landscape Approximation}\label{sec:bid landscape}

For CPM, since the cost of an impression is equal to media cost plus fees that are constant (or of constant percentage), it roughly equals to a multiple of the bid price. Hence, we can use a model as simple as 
\begin{equation}\label{eq: cpi_reg}
    CPM_i^k=a+b\cdot f_i^k                                     
\end{equation}
Note that the actual bid price affected by all bid factors $f_i^m, m\neq k$. But in practice we found this approximation to be sufficiently accurate. 

We can also come up with more complicated models to capture the interactive effect of bid factors from different dimensions on the impression cost and volume. For example, we also tested the following model for impression volume:
\begin{equation}\label{eq: volume_marginal}
n_i^k=g\left(f_1^1,\ldots,f_I^K\right) 
=a^k+\left(\beta_i^k\cdot f_i^k\right)\cdot\left(\prod_{d\neq k}^{K}\sum_{j=1}^{I}\beta_j^d\cdot f_j^d\right)    
\end{equation}
Here, we assume the impression volume from group $i$ of dimension $k$ is determined by the product of two terms (plus intercept). The first term is essentially its own bid factor $f_i^k$. The second term is the product of weighted sums of all other bid factors from other dimensions. One can easily see that $n_i^k$ is monotonically increasing w.r.t any bid factor from other dimensions marginally. 

\begin{figure}
    \centering
    \includegraphics[width=\linewidth]{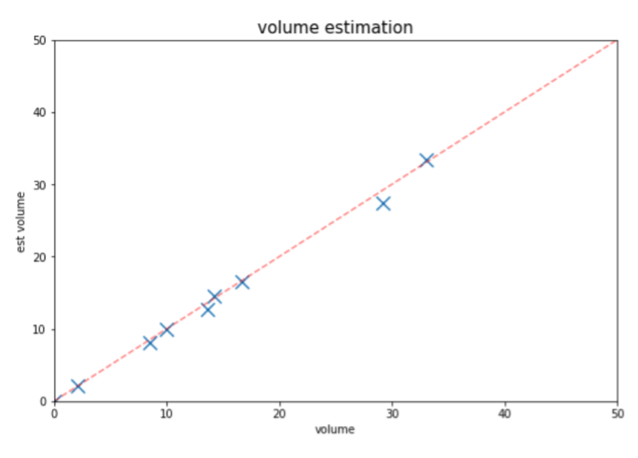}
    \caption{Estimated impression volume based on the model descriped in Eq.~\eqref{eq: volume_marginal} vs. observed impression volume for a given day in an online test campaign. y-axis is the estimated volume and x-axis is the observed. Both values have been scaled.}
    \label{fig:volum_plot}
\end{figure}

The models can be retrained frequently depending on the refresh frequency of the impression log. If impression log refreshes daily, we can update the parameters $a, b$ in Eq.~\eqref{eq: cpi_reg} or ${a^k}, \{\beta_i^k\}_{i\in[I], k\in[K]}$ in Eq. ~\eqref{eq: volume_marginal} on a daily basis. Figure ~\ref{fig:volum_plot} compares the predicted next-day impression volume for each group of all dimensions based on the model described in Eq.~\eqref{eq: volume_marginal} and the real impression volume observed for one of our test campaigns.

\section{Bid Dimension Selection} \label{sec:dimension}
The effectiveness of the optimization system depends heavily on the validity of its assumptions. In Section ~\ref{sec:bid landscape}, we demonstrated the accuracy of the bid landscape approximation in real online campaigns. But we also need to verify our assumptions on RPM, which states that the RPM for each group under a dimension is stable and independent from bid price. 
The behaviors of dimensional RPM depends on how we define those dimensions. Ideally, the dimension and its groups should be chosen in such a way that the difference in RPM from different groups under a dimension is maximized. In other words, the dimension of choice should be indicative of revenue and its groups should capture the relationship between this dimension and revenue. For example, we can choose zip code as a dimension and group them into different buckets. The purchasing behavior of people from different geographical regions tend to differ; therefore, these zip buckets will have different ROAS potential for us to exploit. Hour of day is another feature of the impression supply during prime hours is significantly higher than other periods, which leads to cost and profitability differentiation.  Other possible candidates for dimensions include OS type, device type, day of week, etc.

\subsection{Equal-volume with Ranked RPM}

No matter what criteria we use to define segments or groups under a dimension, we need to make sure there are enough impressions for us to observe stable revenue signals. Thus, one natural option is group directly by impression volume. But randomly putting impressions into groups such that the groups have roughly the same volume does not guarantee separations of the groups in terms of RPM. Thus, we propose a two-step heuristic to define the groups. 
\begin{enumerate}
    \item First, we sort all unique dimensional features values (unique zip codes, websites, hours, etc. observed in log so far) based on their accumulative RPM. 
    \item Second, we sequentially assign group labels to each value, from the one with the lowest RPM to the highest. We will move to the next group label once the current one has enough impression volume based on historical data. 
\end{enumerate}

Taking zip code as an example, in our implementation, we calculate the total impressions for each unique zip code and sort them by their accumulative RPM so far in ascending order. Then, we put the top zip code into group 1. Depending on the number of groups we set, we iteratively put the next zip code into a bucket so that at the end, all buckets have roughly the same number of impressions. Other than using RPM has the metric for sorting the values, we also experimented using conversion counts or order counts. Using conversion counts to sort previous the groups from being skewed by high-value purchasers when the total number of impression is low. In practice, we also suggest setting thresholds on either volume or orders before estimating the RPM. The system can wait until there is enough sales signal to trigger the group creation and dimensional adjustments. 

More grouping methods can be developed based on the specific dimensions we choose. For example, zip codes are hierarchical and define a natural geological clustering; thus we can group them by the first few digits. Website can also be grouped based on their categories \cite{comscore}. Hours of day and week can be easily grouped into prime and low-velocity periods. Yet, the question remains: how do we decide which dimension to include in the bid adjustments and how many groups should each dimension have? To assess the ``quality'' of the dimensions we choose and the groups defined, we also propose several evaluation criterion. 

\subsection{Bid Dimension and Group Evaluation}\label{sec: quality}
\subsubsection{Evaluation by RPM Distances}
\begin{figure}
    \centering
    \includegraphics[width=\linewidth]{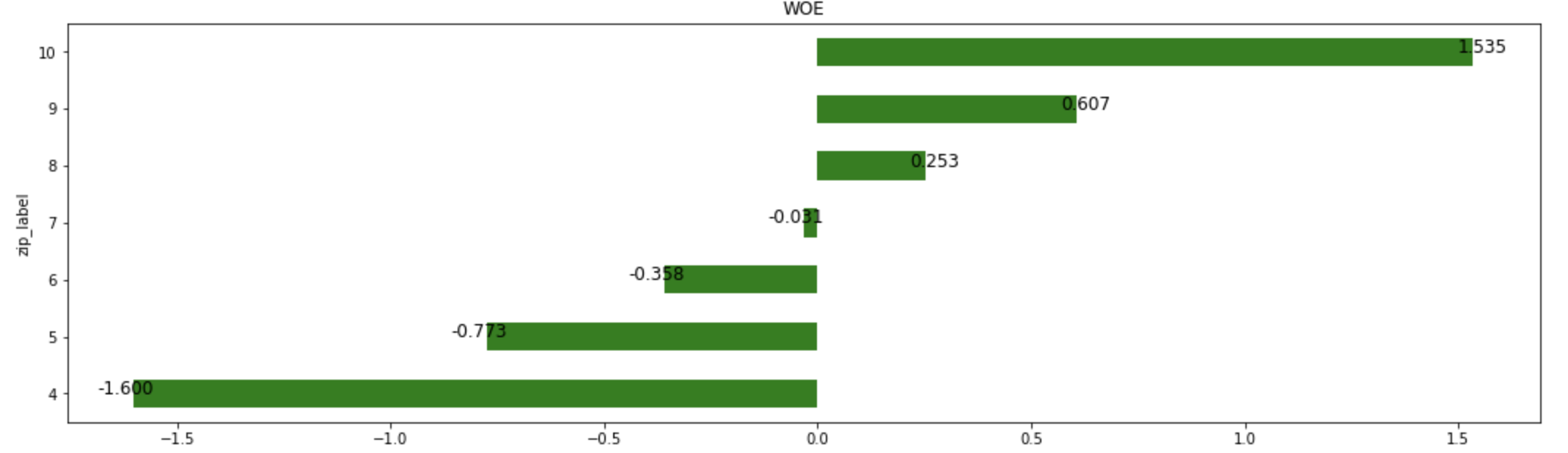}
    \caption{Weight of Evidence plot for groups created on dimension $\textit{zip code}$. 7 groups are created in total.}
    \label{fig:woe_zip}
\end{figure}

\begin{figure}
    \centering
    \includegraphics[width=\linewidth]{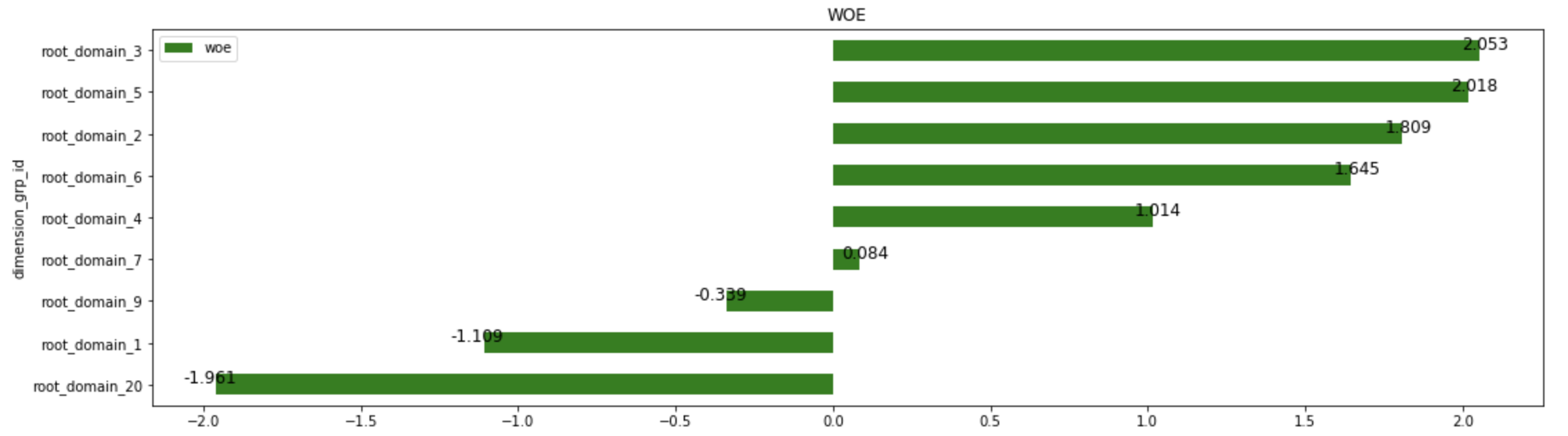}
    \caption{Weight of Evidence plot for groups created on dimension $\textit{website}$. 9 groups are created in total.}
    \label{fig:woe_site}
\end{figure}

A straightforward way of determining the \textit{separation} of different groups in terms of RPM is comparing their daily accumulative RPM numbers. This daily accumulative defines a time series of RPMs and if the difference between the two time series from two different groups is large, we can say the groups are well-separated and vice versa. To quantitatively measure the difference, we propose the following heuristic measure. Suppose we have $N$ groups under a dimension, we first compute the daily accumulative RPM for each group. Then, we use the absolute value of the sum of daily differences to represent the distance between two groups, as shown in Eq.~\eqref{eq: dist}. We also define a metric $r(a,b)$ as in Eq.~\eqref{eq: r} to measure the number of cross-overs between the two time series; the fewer the cross-overs, the higher $r$ will be, and thus, the bigger the final distance will be. 

\begin{equation}\label{eq: r}
    r(a, b) = \frac{\sum_{t=1}^T \mathbf{1}\left(RPM_{\text{grp}\ a}^t > RPM_{\text{grp}\ b}^t\right)+1}{\sum_{t=1}^T \mathbf{1}\left(RPM_{\text{grp}\ a}^t \leq RPM_{\text{grp}\ b}^t\right) + 1} 
\end{equation}

\begin{equation}\label{eq: dist}
\begin{aligned}
    \text{pairwise\_distance} (a, b) &= \biggl| \sum_{t=1}^T \left(RPM_{\text{grp}\ a}^t - RPM_{\text{grp}\ b}^t\right)\biggr| \\ &+ \lambda \cdot \frac{\max\ \{r, 1/r\}}{T} 
\end{aligned}
\end{equation}
In the end, we can use the median or average distances among all pairs of groups to quantify the quality of the set of groups defined. Figure ~\ref{fig:woe_zip} and ~\ref{fig:woe_site} show the WOE values of different groups created based on dimension zip code and website. The large variance in the WOE values among the groups indicate a high predicative power of the groups on RPM.

\subsubsection{Evaluation by Information Value Criteria}

Another method we implemented is based on information theory.  Intuitively, we want the revenue distributions from different groups of a dimension to be as different as possible so that different bid adjustments can be applied. From information theory we have a metric to measure exactly this difference – information value (IV). The higher IV a variable has, the more predictive power it has towards the target variable. IV is defined as the weighted sum of weight of evidence (WOE). [3] We used a slightly modified version of WOE to accommodate continuous features. $\%\text{conversions}_i$ is defined as the percentage of converted impressions in group $i$ of a single dimensions. $\%\text{impressions}_i$ is the percentage of impressions from group $i$ out of all impressions. Given a dimension and its groups, we calculate the percentage of revenue, impressions falling into each group w.r.t. the total revenue and total impression volume, respectively. The modified WOE is defined as the log ratio between the two. 

\begin{equation}
    \text{Modified WOE}_i = \left[\log\left(\frac{\%\text{conversions}_i}{\%\text{impressions}_i}\right)\right]\times 100\%.
\end{equation}

\begin{equation}
\begin{aligned}
    IV &= \sum_{i=1}^n \left(\%\text{conversions}_i - \%\text{impressions}_i\right)\times \\
    &\quad \log\left(\frac{\%\text{conversions}_i}{\%\text{impressions}_i}\right).
\end{aligned}
\end{equation}

\subsubsection{Mutual Information Value Criteria}

The two criterion introduced above are mostly for evaluating groups of a single dimension. To evaluate multiple dimensions, we propose the following metric based on the Mutual Information \cite{vergara2014review}. Let $Y\in \{0,1\}$ be the conversion flag variable of an impression, $g_i$ be the group label variable of dimension $i$, $p(\cdot)$ be the empirical density function. 
\begin{equation}
    MI(g_i, g_j, Y) = \sum_{(i,j)\in (g_i, g_j)} \sum_{y\in Y} p(i, j, y)\cdot\log\left(\frac{p(i,j,y)}{p(i,j)\cdot p(y)}\right)
\end{equation}
Here $p(i,j)$ is the density function of the joint distribution of impressions over the pairs of groups from two dimensions. Likewise, $p(i,j,y)$ is the density function of the joint distribution of $(g_i, g_j, Y)$ The higher $MI(g_i, g_j, Y)$ is, the more information the group pair $(g_i, g_j)$ contains about the conversion signal.

\section{Online Test Results}\label{sec:results}
\begin{figure}
    \centering
    \includegraphics[width=\linewidth]{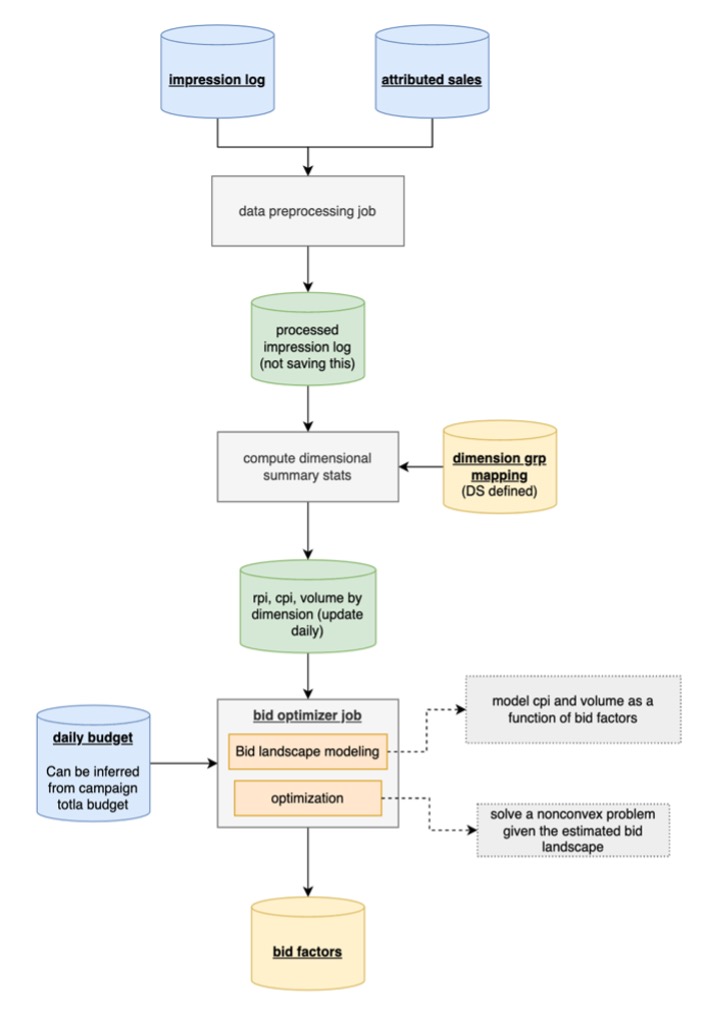}
    \caption{Bid optimization system flowchart based on dimensional bidding. Overall, it is comprised of a data processing job and an optimization job. The \texttt{impression log} is owned and refreshed by a 3P DSP while the \texttt{attributed sales} data is created by the retailer.}
    \label{fig:design}
\end{figure}

We ran a proof-of-concept test campaign together with a control campaign to evaluate the bid optimization framework. Both a test and a control campaign were set up using the same targeting criterion and launched simultaneously for the same flight duration of one month. The targeted audiences were split randomly into the test and control groups. The campaign budgets for the test and control are also the same. The overall system operates as described in Figure ~\ref{fig:design}.

 Based on the data we collected during the first week of the campaigns, we defined our groups based on two dimensions: designated-marketing-area (DMA) and site. DMAs were split into 5 groups while the sites into 4. We used selection by volume method described in Section 3.1. Throughout the campaign period, we kept track of the accumulative RPM values for each group and plot them in Figure 7. We can see clear separations among the curves indicting distinctive per-impression revenue patterns. The increasing pattern is due to a 30-day sales attribution window. 


\begin{table}[]
    \centering
    \begin{tabular}{c|c |c |c| c| c}
    \hline
         & Cost & Sales & ROAS& Trans \%& eCPM \\
    \hline
        Control & 1.15K & 13.64K & 11.86 & 11.12\% & 2.88\\
    \hline
    Test & 1.15K & 13.94K & 12.12 & 11.23\% & 2.84\\
    \hline
    \end{tabular}
    \caption{Performance summary of online test and control campaigns. Costs are the same as the two campaigns have equal amount of budget. Sales are attributed based on a 30-day attribution window. Trans \% is equal to the total amount of orders divided by total impression volume. eCPM is the effective Cost Per Mille impressions. While costs are equal, the test campaign achieved higher revenue, ROAS, transaction rate with lower eCPM. For confidentiality reason, all numbers have been rescaled.}
    \label{tab:test_campaign}
\end{table}

\begin{figure}
    \centering
    \includegraphics[width=\linewidth]{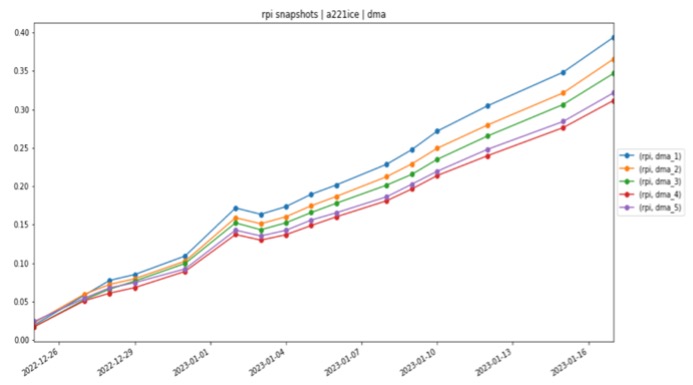}
    \caption{Accumulative daily RPM plot for 5 groups created based on dimension $\textit{designated market area (DMA)}$. The separation between two RPM curves indicate distinct profitability of impression from the two DMA groups. }
    \label{fig:rpi}
\end{figure}

The bid adjustments were performed roughly every other day. At the end of the campaign, we measured the total cost, total revenue, ROAS, and eCPM for both the test and control campaigns. In Table ~\ref{tab:test_campaign}, we can see the ROAS was increase by \$0.26, or around 2\%. The conversion rate was also increased slightly from 11.12\% to 11.23\% overall.  Despite having lower eCPM, the test campaign achieved higher total sales. 
\section{Conclusions and Future Work}\label{sec:summary}
In this work we proposed a bid optimization framework for offsite ads campaigns on 3P DSPs. The framework applies bid adjustments to the base bid price based on the different segments of the impression universe. Instead of treating each segment as independent, we assign bid adjustment on marginal dimensions to avoid parameter explosion and the curse of dimensionality. By translating the problem into a constraint non-convex optimization problem, we can find locally optimal bid factors to maximize daily revenue. 

The performance of the initial version of the optimization framework was tested in production environment based on randomized experiments. We made bid adjustments every other day over the flight of the campaign and achieved higher ROAS than the control group with statistical significance. The online experiment was conduct in a strict A/B testing fashion. 

However, there are several issues yet to be resolved. For example, the current way of defining dimension and groups is very crude as it does not guarantee maximum and stable separations in terms of revenue per impression or conversion rate among the bid dimensions and groups. For the next step, we will explore other ways, such as combining prior knowledge of campaigns under the same category with historical data. In our initial experiment with the method, we already were able to identify that zip code-based bid groups are more indicative of revenue behavior than features such as hour of day using information value. We believe later when we also incorporate interactive effect metrics for measuring pairs of dimensions, we will identify the bid dimensions and groups more efficiently. 

\begin{acks}
Special thanks to Yuan Feng, Erdi Gao, Biyi Fang, Chaowen Zheng, Kyle Mauseth, etc. for inspiring discussions and the folks in the engineering team at Walmart Global Tech for their support.
\end{acks}


\bibliographystyle{ACM-Reference-Format}
\nocite{*} 
\bibliography{paper_sections/references}

\appendix

\end{document}